\documentclass{article}

\usepackage{graphicx, cite, comment, authblk, amsmath, caption, enumitem}
\title{Detecting Crypto Pump-and-Dump Schemes: A Thresholding-Based Approach to Handling Market Noise}
\author{Mahya Karbalaii}
\affil{\small{LUISS Data Lab, LUISS Guido Carli, Viale Pola, 12, 00198 Roma, Italy\\ mkarbalaii@luiss.it}}
\date{February 2025}

\begin{document}
\maketitle

\begin{abstract}
We propose a simple yet robust unsupervised model to detect pump-and-dump events on tokens listed on the Poloniex Exchange platform. By combining threshold-based criteria with exponentially weighted moving averages (EWMA) and volatility measures, our approach effectively distinguishes genuine anomalies from minor trading fluctuations, even for tokens with low liquidity and prolonged inactivity. These characteristics present a unique challenge, as standard anomaly-detection methods often over-flag negligible volume spikes. Our framework overcomes this issue by tailoring both price and volume thresholds to the specific trading patterns observed, resulting in a model that balances high true-positive detection with minimal noise.
\end{abstract}


\section{Introduction}
As cryptocurrency markets continue to gain popularity, an increasing number of individuals are eager to invest in these emerging financial ecosystems, seeking to capitalize on the opportunities presented by blockchain technology. Beside being highly attractive from investment point of view, these digital currencies facilitate money transactions without the need of the approval by authorities such as banks. These characteristics have attracted millions of people to join this market. According to Statista the number of verified cryptocurrency holders has surpassed 600 million people and trend is definitely growing with no-sign of relentlessness. The total market capitalization of cryptocurrency has reached 4 trillion USD at the end of 2024 according to the Economist.

However, as with any rapidly expanding market, the influx of new investors also attracts malicious actors looking to exploit market inefficiencies and the inexperience of retail traders. One of the most prominent fraudulent schemes observed in these markets is the "pump and dump" strategy, a form of market manipulation that artificially inflates asset prices before a coordinated sell-off. The technical complexities in this market makes it also very difficult to define regulations and even harder to enforce such regulations. As a result, unlike the traditional trading platforms and markets, the possibilities to control the crypto-tradings remain questionable.

While such market manipulations have a long history \cite{lee2008pump,lee2010information, cumming2015fraud, johnson1999market}, the presence of the authorities made it possible to limit them. The pump-and-dump (P\&D) scheme \cite{crypt_schemes, anatomy_pump, crypto_pump_scheme}, which is at the focus of this article, has existed for over a century and was notably prevalent in traditional stock markets before strict regulations were enforced. During the dot-com bubble in the late 1990s and early 2000s, many brokerage firms engaged in such manipulative practices, promoting low-value “penny stocks” to unsuspecting investors, only to abandon them once prices had been artificially driven up. Recognizing the fraudulent nature of this scheme, financial regulatory bodies such as the U.S. Securities and Exchange Commission (SEC) \cite{sec} deemed pump-and-dump operations illegal, introducing strict penalties for those involved.

However, in cryptocurrency markets, regulatory oversight remains inconsistent across different jurisdictions. Many crypto-exchanges operate with limited or no regulatory frameworks, creating an environment where pump-and-dump schemes flourish \cite{gandal2018price}. Unlike traditional financial markets, where strict disclosure requirements and trading surveillance mechanisms help detect manipulative activities, cryptocurrency exchanges often lack these safeguards, allowing bad actors to orchestrate and execute such schemes with relative ease.

Although such phenomenon are gaining attention among the scientists \cite{clough2023edwards, sequence_pump_dump, ml_real_time_pump_dump}, still the research around them remain scarce. The observed patterns in the P\&D events can very well be confronted by anomalous patterns in complex systems. These similarities therefore encourage the use of anomaly detection algorithms to identify such events. 
In the first part of this article, we explain the mechanisms of pump-and-dump schemes in cryptocurrency markets. The second part of this work is concentrated on using unsupervised techniques to identify such events. We compare the results of our models with verified announcements of pump events to evaluate its performance.

\section{Background}
In this section we provide some technical definition for the market manipulation activities, the psychological facts which encourage people to take part in such events and their consequences. We also briefly review the economical impacts of such malicious activities and why they are gaining ground so rapidly in the emerging world of cryptocurrencies.

\subsection{Centralized and Decentralized Exchanges (CEXs and DEXs)}
The execution of pump-and-dump schemes varies depending on the type of cryptocurrency exchange used. Cryptocurrency exchanges can be broadly categorized into centralized exchanges (CEXs) and decentralized exchanges (DEXs), each with distinct characteristics and vulnerabilities.

Centralized exchanges such as Binance, Coinbase, and Kraken operate as registered entities that provide trading platforms, order books, and custodial services for users. CEXs typically:
\begin{itemize}[noitemsep]
    \item Require user identity verification (Know Your Customer, KYC regulations).
    \item Maintain liquidity and regulate order matching.
    \item Facilitate fiat-to-crypto transactions.
    \item Comply with governmental regulations to prevent money laundering and fraud \cite{fatf}.
\end{itemize}

Due to their structured nature and compliance with regulations, CEXs offer some level of protection against fraudulent activities. However, they are also more prone to regulatory scrutiny, which may deter certain investors, and security risks, as they maintain large pools of user funds, making them targets for hacking attacks \cite{crypto_crime}.\newline
Decentralized Exchanges (DEXs), such as Uniswap and PancakeSwap, operate without a central authority, relying on smart contracts to facilitate transactions. Key characteristics include:
\begin{itemize}[noitemsep]
    \item Anonymity: Users do not need to complete KYC procedures.
    \item On-chain trading: All transactions are recorded directly on the blockchain, reducing counter-party risks.
    \item Non-custodial nature: Users retain full control over their private keys and assets.
    \item Limited fiat integration: Only cryptocurrency pairs can be traded, excluding fiat transactions.
\end{itemize}

Because of these characteristics, DEXs are less susceptible to regulatory oversight but are more prone to market manipulation, as there are fewer restrictions on the creation and promotion of tokens, as well as flash loan attacks, where large, temporary capital injections distort prices. Scam projects, particularly with the rise of easily created "meme tokens" that serve as vehicles for fraudulent activities \cite{anatomy_pump, ml_real_time_pump_dump, sequence_pump_dump} are more spread on DEXs.

\subsection{The Future of Cryptocurrency Exchanges and Market Manipulation}
As decentralized finance (DeFi) continues to grow, DEXs are becoming increasingly popular due to their security and resistance to censorship. However, this rise also presents new challenges for market integrity. While CEXs remain vulnerable to internal and external manipulation, DEXs provide an environment where fraudulent schemes can proliferate with minimal oversight. To mitigate these risks, solutions such as, on-chain analytics tools to detect irregular trading patterns, decentralized identity verification mechanisms, stronger regulatory frameworks for DeFi projects are being explored. Future developments will determine whether cryptocurrency exchanges can effectively balance decentralization with investor protection.

\subsection{Pump-and-Dump Process}
A pump-and-dump scheme is a form of market manipulation in which the price of a specific asset is artificially inflated before being rapidly sold off for profit. While this practice is illegal in traditional stock markets due to regulatory protections \cite{sec}, it remains largely unregulated within cryptocurrency markets. This lack of enforcement allows malicious actors to exploit unsuspecting investors who fall victim to misleading information, leading to significant financial losses.
A pump-and-dump event typically unfolds in three distinct phases \cite{anatomy_pump, crypt_schemes, crypto_pump_scheme}: accumulation, pumping, and dumping.

\textbf{Accumulation Phase:}
This initial phase is characterized by a slow and unnoticed accumulation of a target cryptocurrency token by insiders. These insiders, often a small group of organized manipulators, begin purchasing substantial quantities of the token over time. Because these purchases are made gradually, they do not significantly affect market prices or attract attention from other investors.

\textbf{Pumping Phase:}
In this phase, the organizers begin a coordinated effort to inflate the price of the token. This is primarily achieved through widespread marketing campaigns using social media platforms such as Telegram, Discord, Reddit, and Twitter \cite{ante}. These platforms provide anonymity, encrypted messaging, and large audiences, making them ideal for spreading misleading information about the token’s potential. Organizers often:
\begin{itemize}[noitemsep]
    \item Set a specific date and time for the pump.
    \item Instruct subscribers to prepare funds for the event.
    \item Create artificial hype around the token.
    \item Utilize bots or influencers to amplify the message.
\end{itemize}
Importantly, the identity of the target token remains undisclosed to participants until the very last moment to prevent premature trading activity. Once revealed, an influx of buy orders rapidly increases the token’s price, driven by fear of missing out (FOMO) among investors.

\textbf{Dumping Phase:}
The final phase involves the rapid selling of holdings by insiders who accumulated tokens at low prices. As prices peak due to increased demand, these organizers liquidate their positions, generating substantial profits. This triggers a cascading effect:
\begin{itemize}[noitemsep]
    \item Initial investors begin selling to secure profits.
    \item The price starts to decline rapidly.
    \item Panic selling ensues as other participants attempt to exit before further losses.
\end{itemize}
The entire cycle can last anywhere from a few seconds to a few minutes. In many cases, unsuspecting investors are left holding worthless tokens, having purchased them at inflated prices \cite{to_the_moon}.

\section{Unsupervised Anomaly Detection}
In the context of pump-and-dump events, anomalies can be interpreted as sharp spikes in both price and trading volume of a token, which are not observed in the rest of the dataset \cite{ahmed2020unsupervised, chalapathy2019deep}. They are short intervals in normal trading patterns where both price and the volume of an specific token increases and suddenly decreases. However, traditional unsupervised machine learning techniques and models show to be little effective in identifying anomalies in time-series datasets.

In general, anomalies can be categorized into three main types \cite{chandola2009anomaly, song2019conditional}:\newline
\textbf{Point Anomalies:} These occur when an individual data point exhibits behavior significantly different from the rest. For instance, a sudden, extreme price surge in a normally stable asset may be considered a point anomaly.\newline
\textbf{Collective Anomalies:} These arise when a group of data points collectively exhibit anomalous behavior, even if individual points within the group may not appear anomalous on their own. An example would be a series of small price increases over a short period that, when taken together, indicate an ongoing pump scheme.\newline
\textbf{Contextual (Conditional) Anomalies:} These anomalies depend on the surrounding context. A value that is normal in one situation might be considered anomalous in another. For example, an unusually high trading volume at a specific time of day could be normal for a particular token but would be anomalous if it occurred during typically low-activity hours. A real-world analogy would be high temperatures being normal in the summer but considered anomalous in the winter.

\subsection {Thresholding Technique}
One of the first research works using unsupervised anomaly detection approach was deployed by \cite{to_the_moon} to identify pump-and-dump events across various cryptocurrency exchanges.
To gather and construct their dataset, the authors initially identified 24 exchanges. From each exchange, they selected 50 trading pairs (symbols). For each symbol, they obtained historical data of OHLCV (Open, High, Low, Close, Volume) \cite{brown2012candlestick, gandal2018price} for a period of 20 days at hourly intervals. Ultimately, the dataset was compiled using data from five exchanges and 450 symbols, amounting to 480 hourly candles per trading pair.
Since the authors aimed to work with unlabeled training data, they opted for an unsupervised anomaly detection approach. Anomaly detection methods are essentially algorithms designed to identify outliers — data points that deviate significantly from the rest of the dataset due to their distinct behavior.

In their paper, the authors focused on identifying indicators that could help detect anomalies, specifically opting for contextual or conditional anomalies. The goal was to define thresholds for certain key indicators—referred to as indicator variables—that have a direct impact on the detection process. Additionally, they identified environmental variables that indirectly influence the emergence of anomalies.
Their goal was to locate corresponding spikes in both price and volume and establish thresholds that define anomalous behavior. Once these indicators surpass their respective thresholds, an anomaly can be identified.
The authors considered pump-and-dump events as inherently local phenomena. Consequently, they used threshold-based techniques to detect such localized anomalies. To achieve this, they employed a moving window approach, which utilizes a simple moving average of the selected variable. By comparing the moving average within a given window to that of the previous window, they were able to detect deviations from the trend. This method helps capture trends over time rather than focusing on isolated values. The time window used in this approach is commonly referred to as the lag factor.

According to their definition, if the highest price at any given point exceeds the anomaly threshold, that point is classified as anomalous. To compute the price and corresponding volume thresholds, the authors proposed the following methodology:
Given a percentage increase  and a lag factor , the moving average  is calculated as (formula to be inserted later). Similarly, an anomaly threshold is defined for volume.
The final objective is to optimize both price and volume thresholds so that if a point’s corresponding variables surpass both thresholds, it is classified as a local conditional anomaly.

\subsection{Verification of Organized Events}
Starting from the conceptual framework introduced by \cite{to_the_moon}, we have selected a dataset to apply an analogous method. However, given the rapidly evolving nature of the cryptocurrency ecosystem, many aspects have changed since the publication of their paper. For example, one of the principal sources previously used by many research works to find the list of Telegram channels that announce pump-and-dump events is no longer available. This served as a ground-truth reference for verifying actual pump events, is now defunct, necessitating a manual search to identify such channels on Telegram.
Accordingly, we conducted a systematic manual search for various public Telegram channels that regularly broadcast pump-and-dump events. Establishing a reliable ground-truth is essential for verifying both the performance of our model and the validity of our results. The selected channels typically announce approximately one pump event every three to four days and maintain subscriber counts in the range of 70,000 to 80,000.

The organization of these pump events generally follows a structured sequence. Usually four to five days before an event, an administrator—often operating as an automated bot—issues an initial announcement specifying that a particular coin will be pumped at a predetermined date and time. This announcement includes details such as the corresponding exchange and the specific trading pair (i.e., the quote currency), thereby encouraging subscribers to fund their wallets on the indicated exchange with the appropriate assets. Following the initial notification, a series of countdown messages are broadcast over the subsequent days. As the scheduled time approaches, these countdowns transition from daily updates to hourly intervals, and finally to minute-by-minute alerts. At the designated moment, the coin’s name is revealed, prompting a significant portion of the channel’s audience to participate in the pump event. Shortly thereafter, organizers typically publish screenshots demonstrating how the coin’s candlestick patterns have escalated, often showcasing the substantial profits accrued by some participants.
With these ground-truth and verification details established, our analysis has focused on the Poloniex Exchange—a platform that emerged as the most frequently referenced among the Telegram channels surveyed.

\section{Model Setting and Challenges} 
A thorough understanding of the dataset's characteristics is essential for constructing a robust machine learning model. In particular, factors such as statistical properties, distribution, volatility, and noise levels must be carefully considered. These considerations ensure that the model is well-calibrated to capture both broad market trends and the rapid fluctuations characteristic of pump-and-dump events.

\subsection{Dataset}
By concentrating on the Poloniex Exchange, we retrieved a complete list of its markets via its public API. At the time of data collection, Poloniex listed over 1,100 trading pairs. Poloniex is a cryptocurrency centralized exchange (CEX) platform associated with Justin Sun, the founder of Tron. Notably, our examination of Telegram revealed that many of these pairs are quoted against Tether (USDT).

Although Poloniex provides free access to its public API endpoints, it enforces a limit on the number of calls. Consequently, we downloaded the data in segmented chunks to avoid exceeding the exchange’s call restrictions. Our initial dataset comprises detailed information for all registered trading pairs on Poloniex over a six-month period, from 15 August 2024 to 15 February 2025. The features collected for each pair include, but are not limited to, market metrics such as open, high, low, close, and volume data which led to a dataset for 4.3 million data points.

\subsection{Double Conditioning and Denoising}
It is widely recognized that tokens targeted in pump-and-dump events often possess extremely low market capitalizations. However, our investigation indicates that low market capitalization is not the sole distinguishing feature. Many of these tokens also exhibit prolonged periods of inactivity, interspersed with additional lengthy dormant intervals. Although such tokens are generally not de-listed, market participants may engage in minimal trading activity to avoid triggering a de-listing, complicating the detection of anomalous behavior.
Our initial approach to identifying anomalous trades was inspired by the methodology presented in \cite{to_the_moon}. Specifically, we defined a moving average tail for both price and volume to compare the current value of each feature with its preceding moving average. The hypothesis was that a sudden and significant increase in both price and volume might indicate a pump-and-dump event. However, the observed trading patterns required a more nuanced threshold-setting strategy.

\begin{figure}
    \centering
    \includegraphics[width=1.0\linewidth]{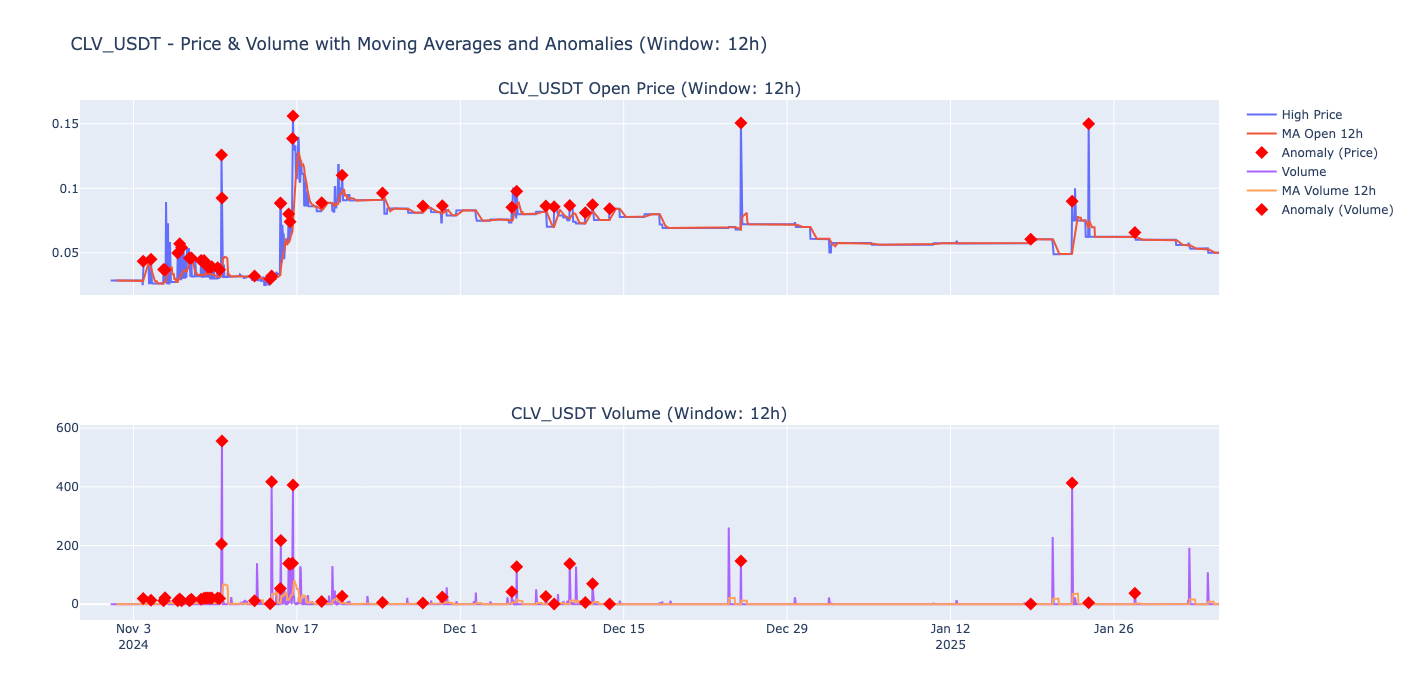}
    \caption{\footnotesize\textit{Relying on straightforward thresholding techniques results in too many anomalous points}}
    \label{fig:noisy_volume}
\end{figure}

A straightforward application of fixed thresholds for price and volume increases resulted in an unrealistically high number of detected events (see Figure \ref{fig:noisy_volume}). Consequently, it became necessary to develop more sophisticated criteria that adapt to the specific characteristics of our dataset. For example, many trading pairs listed on Poloniex exhibit extended inactivity periods during which the OHLC (Open, High, Low, Close) \cite{lu2025hua, Agarwal2024prabha} values remain constant at the last traded level and the trading volume drops to zero. As a result, any subsequent minor increase in volume—even one executed solely to prevent de-listing—may be falsely flagged as anomalous when compared to a zero baseline from the preceding 12 hours.
To address these challenges, we devised a screening method for volume anomalies using a double-condition approach. In this way, we set a minimum threshold which defines the eligibility of an instance to be considered for further anomaly detection or not.\newline
For each conditioning scenario, we calculated the moving average of the open price over the past 12 hours and examined the model under the following five threshold settings: 
\begin{enumerate} 
    \item 90\% threshold on open price and 400\% threshold on volume \label{con:con1} 
    \item 70\% threshold on open price and 300\% threshold on volume \label{con:con2} 
    \item 100\% threshold on high price and 400\% threshold on volume \label{con:con3} 
    \item 90\% threshold on high price and 400\% threshold on volume \label{con:con4} 
    \item 80\% threshold on high price and 300\% threshold on volume \label{con:con5} 
\end{enumerate}

\subsection{Total Volume over 30 Days} 
In this scenario, we consider the total trading volume over the preceding 30 days. For a given instance “a” with trading volume V to be eligible to be considered as anomalous, the following conditions must be satisfied at the same time: 
\begin{equation}
\label{eq:tot_conditions}
\begin{cases}
V > 0.30 \times \text{V}_{tot} \\
V > 0.60 \times V_{\max},
\end{cases}
\end{equation}
\noindent where \( \text{V}_{tot} \) is the total trading volume over the past 30 days, and \( V_{\max} \) is the maximum trading volume recorded over the same period.

\begin{figure}
    \centering
    \includegraphics[width=1.0\linewidth]{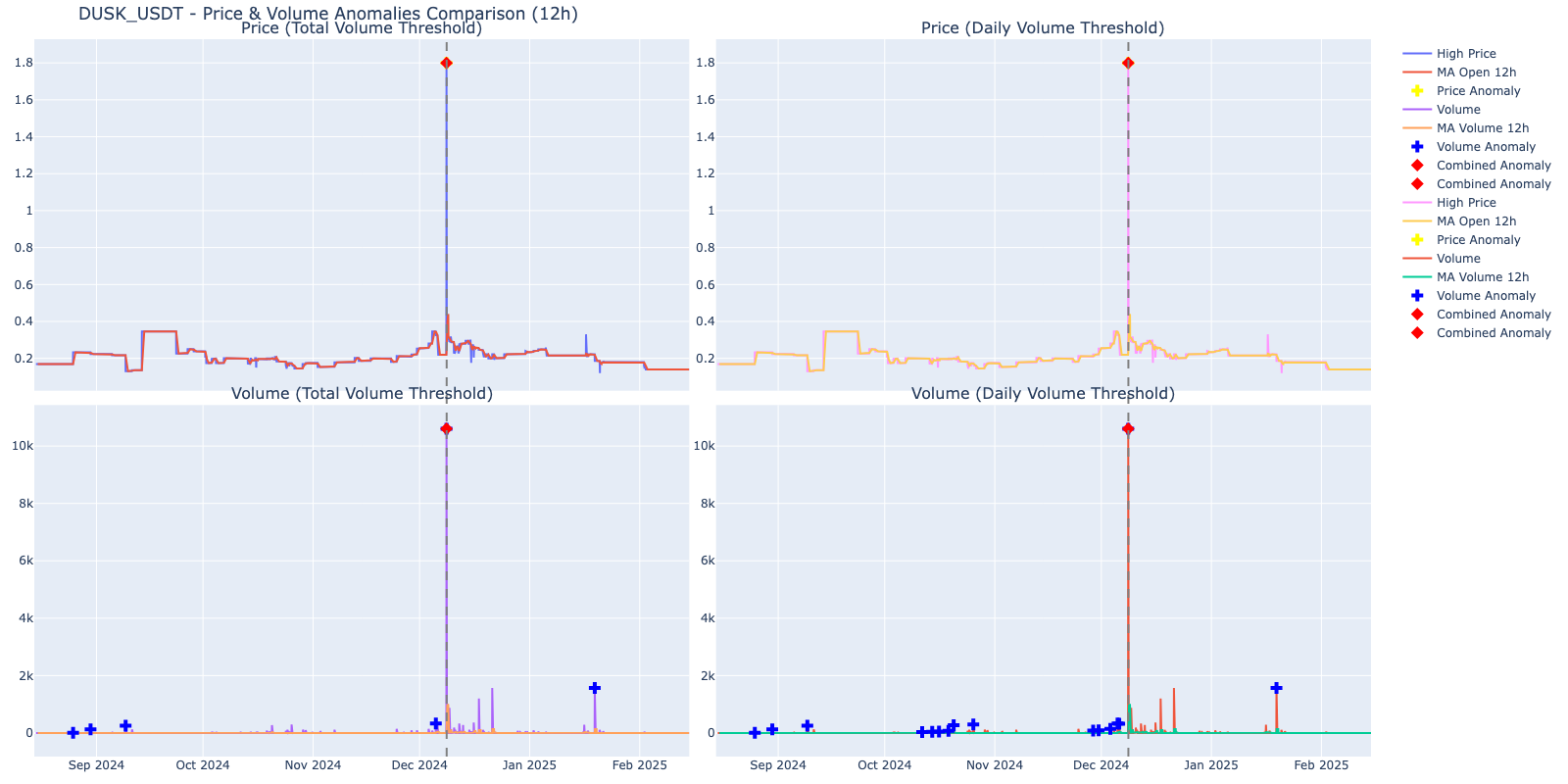}
    \caption{\footnotesize\textit{Double conditioning the volume, filters minor and irrelevant spikes out and the model becomes able to identify the pump events correctly.}}
    \label{fig:total_vol}
\end{figure}

Under these conditions, we assume that in a pump event at least one-third of the monthly trading volume occurs at the pump event. Moreover, since multiple anomalies may be observed, the volume spike must also reach at least 60\% of the highest volume observed over the 30-day window. This double-condition approach effectively filters out minor volume increases following prolonged inactivity.
After filtering out minor spikes, we apply the original threshold criteria (Threshold Settings \ref{con:con1}--\ref{con:con5}) to both volume and price to identify pump events. Recognizing that pump-and-dump events are characterized by simultaneous spikes in both price and volume, we extended our analysis to detect significant price jumps (Figure \ref{fig:total_vol}).\newline
Given that during inactivity the OHLC values remain static at the previous trade’s level, we imposed the price thresholds as specified in the settings. The detected events are then validated against verified pump announcements. Table~\ref{table:tot-vol-30} presents the results of each setting in identifying volume and price anomalies—and their concurrence as “combined anomalies.” The outcomes are tested against confirmed pump event dates and times gathered from Telegram channels, where “true positives” indicate correctly identified events and “missed events” denote confirmed events not flagged as combined anomalies.
\begin{table}[ht] 
\begin{center}
\caption{Model Setting Performance on Total Volume of 30-days}
\label{table:tot-vol-30}
\begin{tabular}{|l|l|l|l|l|l|r|}
\hline
Thresholds & Vol Ano. & Price Ano. & Combined Ano. & True Pos. & Missed\\ \hline
Setting 1 & 4395 & 2324& 194 & 5 & 34\\
Setting 2 & 4402 & 2830& 205 & 5 & 34\\
Setting 3 & 4408 & 2250& 317 & 26 &14\\
Setting 4 & 4408 & 2479& 335 & 27 & 13\\
Setting 5 & 4408 & 2704 & 351& 26 &14\\ \hline
\end{tabular}
\end{center}
\end{table}

\subsection{Average Daily Volume over Past 30 Days} 
While the initial double-condition criteria for volume anomalies were highly effective for tokens with extended inactivity periods, they were less effective for trading pairs exhibiting more regular activity over the preceding 30 days. When considering the total 30-day volume, one might expect that the trade volume during a pump event would represent approximately 30\% of the monthly volume; however, this assumption complicates anomaly detection when volume is more uniformly distributed over time.
\begin{figure}
    \centering
    \includegraphics[width=1.0\linewidth]{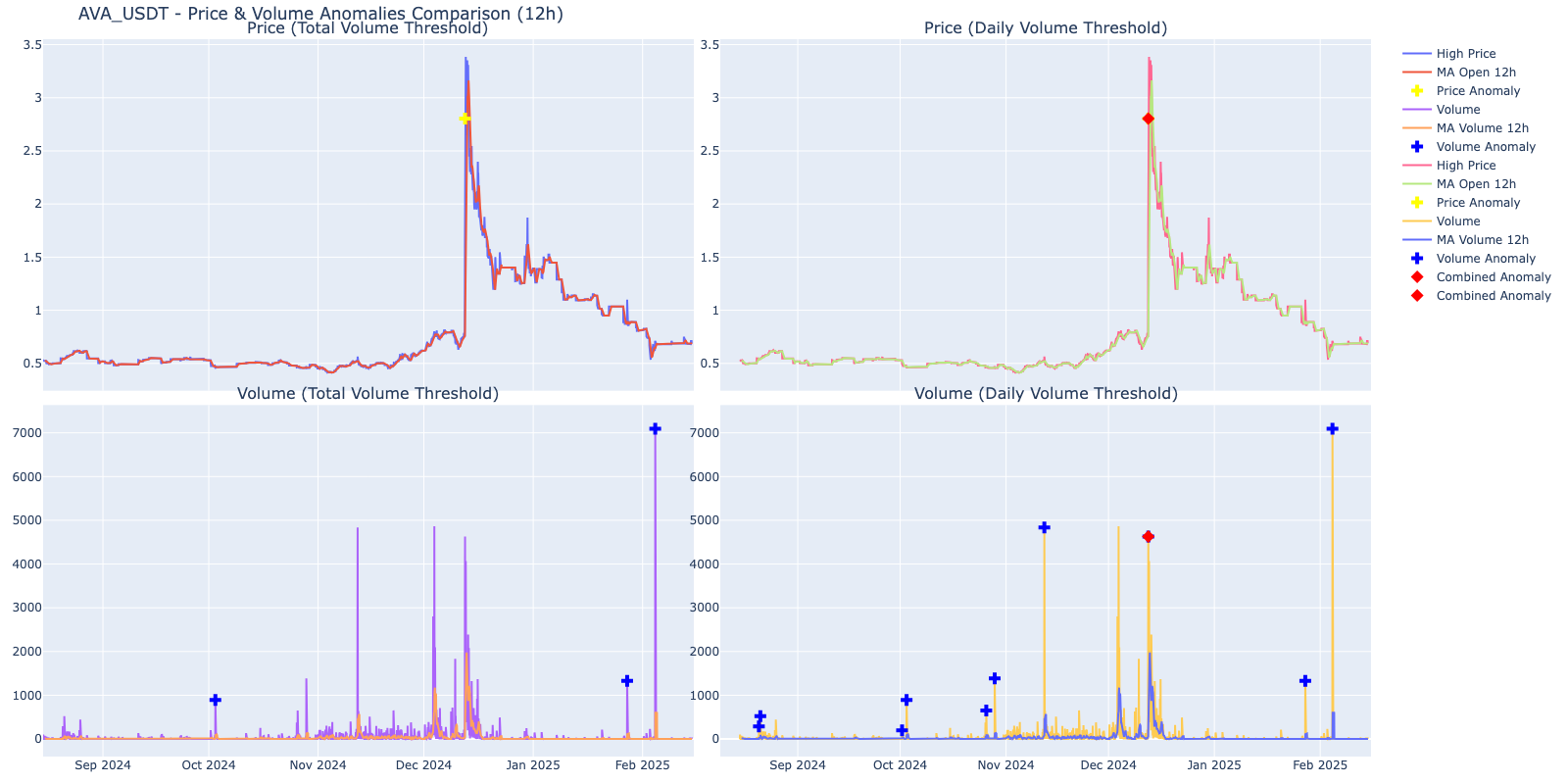}
    \caption{\footnotesize\textit{For tokens with more regular trading volume, the model needs to consider its distribution. Left graphs show how the pump event was missed by considering the Total Volume over the past 30 days, while the right graphs, using the average daily volume show how the final anomalous event was correctly identified. }}
    \label{fig:distribution}
\end{figure}

To capture anomalies in such scenarios, we modified our double-condition approach to account for the distribution of volume on an average daily basis over the past 30 days. Under this revised scheme, a given instance “a” with volume V is considered eligible for anomaly if: 
\begin{equation}
\label{eq:avg_conditions}
\begin{cases}
V > 0.70 \times \text{V}_{avg} \\
V > 0.60 \times V_{\max},
\end{cases}
\end{equation}
\noindent where \( \text{V}_{avg} \) is the average daily trading volume over the past 30 days, and \( V_{\max} \) is the maximum trading volume recorded over the past 30 days.

Table~\ref{table:avg-daily-vol} summarizes the results obtained from all five settings under these new conditions. Notably, Setting No. 4 demonstrates the best performance. An additional observation is that the pattern of true positives and missed events differs between the two conditioning approaches; although overall performance may initially seem reduced, a closer examination across different symbols reveals that the strength of the conditions depends on the specific characteristics of the tokens (e.g., the presence of large, sparse volume intervals).
\begin{table}[ht] 
\begin{center}
\caption{Model Setting Performance on Average Daily Volume}
\label{table:avg-daily-vol}
\begin{tabular}{|l|l|l|l|l|l|r|}
\hline
Thresholds & Vol Ano. & Price Ano. & Combined Ano. & True Pos. & Missed\\ \hline
Setting 1 & 8459 & 2324& 235 & 6 & 30\\
Setting 2 & 8472 & 2830& 257 & 6 & 30\\
Setting 3 & 8479 & 2250& 421 & 24 &16\\
Setting 4 & 8479 & 2479& 451 & 26 & 14\\
Setting 5 & 8480 & 2704 & 477& 24 &16\\ \hline
\end{tabular}
\end{center}
\end{table}

\subsection{Exponential Weighted Moving Average} 
To develop a more adaptable model that effectively captures both categories of tokens, we further refined the conditions by incorporating an exponentially weighted moving average (EWMA) of trading volume. This approach emphasizes recent trading activity. Under the revised criteria, a given instance “a” with volume V is considered eligible for anomaly if: 
\begin{equation}
\label{eq:ewma_conditions}
\begin{cases}
V > 0.70 \times \text{EWMA}_{d} \\
V > 0.60 \times V_{\max},
\end{cases}
\end{equation}

\noindent where \( \text{EWMA}_{d} \) is the exponentially weighted moving average of the past d days, and \( V_{\max} \) is the maximum trading volume recorded over the past 30 days.

As shown in Table~\ref{table:ewma-vol}, testing these revised conditions resulted in a substantial improvement in the model’s capacity to identify true anomalies. In particular, when considering open prices, the number of correctly identified examples doubled respect to the first conditioning method in \ref{eq:tot_conditions}, with additional improvements observed across other settings. Consequently, when these refined volume conditions are combined with the price anomaly criterion, the model successfully identifies multiple pump-and-dump events while minimizing the detection of non-significant trading activities.

\begin{table}[ht] 
\begin{center}
\caption{Model Setting Performance on EWMA with 10 days time span}
\label{table:ewma-vol}
\begin{tabular}{|l|l|l|l|l|l|r|}
\hline
Thresholds & Vol Ano. & Price Ano. & Combined Ano. & True Pos. & Missed\\ \hline
Setting 1 & 7519 & 2324& 248 & 11 & 27\\
Setting 2 & 7533 & 2830& 260 & 11 & 27\\
Setting 3 & 7541 & 2250& 384 & 27 & 13\\
Setting 4 & 7541 & 2479& 415 & 29 & 11\\
Setting 5 & 7541 & 2704 & 440 & 27 &13\\ \hline
\end{tabular}
\end{center}
\end{table}

\subsection{Combined Double-Conditioning and Volatility} 
Finally, to assess the impact of market volatility on our model’s performance, we augmented the volume conditions by incorporating a volatility term. 
For a given instance “a” with volume V, the instance is eligible for anomaly if the following conditions are met:

\begin{equation}
\label{eq:volat_conditions}
\begin{cases}
V > 0.70 \times \text{EWMA}_{d} + \alpha \times \sigma_{\text{daily}}, \\
V > 0.60 \times V_{\max},
\end{cases}
\end{equation}

\noindent where \( \text{EWMA}_{d} \) is the exponentially weighted moving average of the past d days, \( \sigma_{\text{daily}} \) is the daily average standard deviation, \( \alpha \) is a tuning parameter, and \( V_{\max} \) is the maximum trading volume recorded over the past month.
The parameter $\alpha$ adjusts the sensitivity of the model to sudden changes; higher values of $\alpha$ smooth the model’s response by making the threshold more responsive to overall volatility, and vice versa. Table~\ref{table:volatility-vol} presents the model performance for the five threshold settings with $\alpha$ = 2 and d = 10. 
\begin{table}[ht]
\begin{center}
\caption{Model Setting Performance on Volatility with 10 days time span}
\label{table:volatility-vol}
\begin{tabular}{|l|l|l|l|l|l|}
\hline
Thresholds & Vol Ano. & Price Ano. & Combined Ano. & True Pos. & Missed\\ \hline
Setting 1 & 2588 & 2324& 127 & 6 & 30\\
Setting 2 & 2593 & 2830& 131 & 6 & 30\\
Setting 3 & 2597 & 2250& 185 & 19 & 21\\
Setting 4 & 2597 & 2479& 199 & 21 & 19\\
Setting 5 & 2597 & 2704 & 209 & 19 &21\\ \hline
\end{tabular}
\end{center}
\end{table}
While changing $\alpha$ had nearly no effect on the model’s performance, extending the time span of the EWMA from 10 days to 20 days produced a notable impact, both under EWMA-only conditioning and in combination with volatility. Table~\ref{table:volat-ewma} presents the results for Settings\ref{con:con1}, \ref{con:con3}, and \ref{con:con4} according to Equations\ref{eq:ewma_conditions} and \ref{eq:volat_conditions}.

\begin{table}[ht] 
\begin{center}
\caption{Model Setting Performance on Volatility with 20 days time span}
\label{table:volat-ewma}
\captionsetup{skip=10pt}
\begin{tabular}{|l|l|l|l|l|l|}
\hline
  & Setting 1  & Setting 3 & Setting 4 \\ \hline
EWMA Combined Ano. &220 & 382 & 411 \\
EWMA True Positives & 9 & 25 & 26 \\
EWMA Missed Events & 31 & 15 & 14 \\
Volatility Combined Ano. & 164 & 230& 250 \\
Volatility True Positives & 9 & 23 & 25\\ 
Volatility Missed Events & 31 & 17 & 15 \\ \hline
\end{tabular}
\end{center}
\end{table}
\begin{figure}
    \centering
    \includegraphics[width=1.0\linewidth]{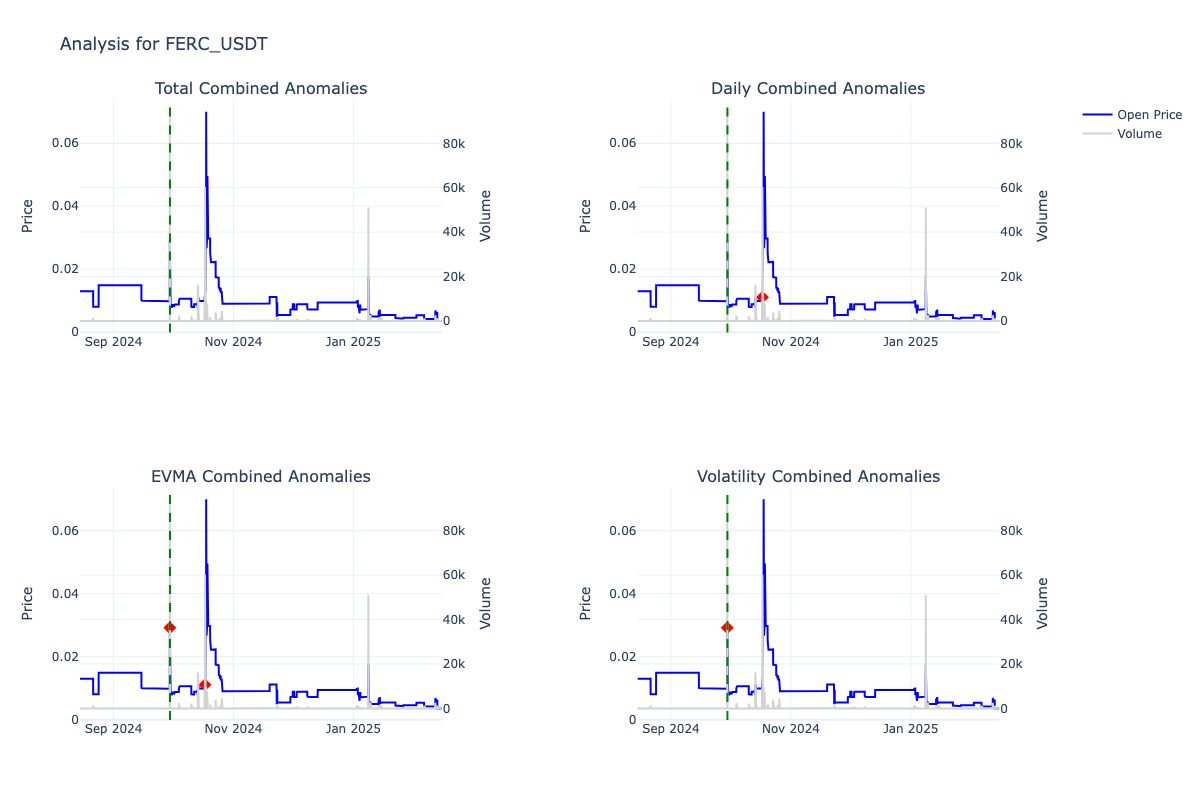}
    \caption{\footnotesize\textit{Comparison of the four conditioning methods: EWMA and EWMA \& Volatility prove to be the best performing choices while both "Total 30-day Volume" and "average Daily" have failed in identifying the verified pump-event. Adding volatility factor to EWMA improves the performance further to identify the only correct pump event. Dashed green line refers to the verified pump event.}}
    \label{fig:best_setting}
\end{figure}
Although thresholding on the open price proved less effective for anomaly detection across all variations, it did show improvements when the EWMA time span was increased from 10 to 20 days. A key observation is the significant improvement of the model when the EWMA is combined with volatility over a 20-day window. This finding aligns with expectations: a 10-day EWMA places greater weight on recent data, making it highly responsive to sudden price or volume spikes. In contrast, a 20-day EWMA smooths out short-term fluctuations, thereby reducing sensitivity to minor anomalies. Although this smoothing might cause the model to miss smaller but potentially meaningful spikes, it is advantageous in filtering out unwanted noise.
Furthermore, it is reasonable to expect that combining EWMA with volatility over a longer time span would enhance performance. Since volatility is typically higher over shorter intervals, transient fluctuations dominate the volatility measure for a 10-day window. Extending the EWMA period to 20 days averages out these transient effects, reducing overall volatility and making spikes appear less dramatic relative to the baseline. Once these smaller spikes are filtered out, the model more accurately detects genuine pump events, thus increasing the number of true positives.

\subsection{Best Setting and Conditioning}
As observed, the best performance on our dataset was achieved with a 90\% increase in the High price—relative to the 12-hour moving average of the open price—combined with a 400\% increase in volume. Using this threshold configuration, we compared various minimum volume thresholds to determine the most effective double-conditioning approach for filtering out minor spikes. Table~\ref{table:final_results} presents the performance of each test under this configuration.
\begin{table}[ht] 
\begin{center}
\caption{Model Setting Performance on Volatility with 20 days time span}
\label{table:final_results}
\captionsetup{skip=10pt}
\begin{tabular}{|l|l|l|l|l|l|}
\hline
 Model & Recall  & F1 Score & Precision \\ \hline
30-day Total Volume &0.65 & 0.68 & 0.71 \\
Average Daily Volume & 0.62 & 0.65 & 0.65 \\
EWMA & 0.65 & 0.65 & 0.65 \\
EWMA-Volatility & 0.62 & 0.71& 0.84 \\ \hline
\end{tabular}
\end{center}
\end{table}

We can therefore, conclude that for the characteristics of our dataset, as the goal of this model is identifying the true pump-events without becoming overwhelmed by other spikes, the combination of EWMA and Volatility on a time span of 20 days works best to filter the irrelevant volume spikes and identifying the most number of pump events (Figure \ref{fig:best_setting}).

\section{Conclusion}
In light of our findings, the adoption of a 20-day EWMA combined with volatility filtering proves to be the most robust strategy for detecting pump-and-dump events in our dataset. By extending the EWMA time span, we achieve a smoother representation of trading volume and price movements, thereby reducing the influence of transient fluctuations that often lead to false positives. This approach is especially valuable for tokens characterized by extended periods of inactivity, where even minor volume upticks can otherwise be misclassified as anomalies.

Moreover, incorporating volatility into the model further refines the detection thresholds. As volatility over shorter time windows tends to be higher, smaller but frequent fluctuations dominate the baseline and inflate the rate of detected anomalies. The longer 20-day period averages out these fluctuations, reducing the baseline volatility and making significant spikes more distinguishable. Consequently, the combination of EWMA and volatility filtering in this extended time-frame strikes a balance between sensitivity (i.e., capturing genuine pump events) and specificity (i.e., filtering out non-relevant spikes).
Finally, although we observed that thresholding on high price alone offers limited effectiveness, it still benefits from a longer EWMA window. The best overall performance was achieved by combining a 90\% increase in High price—relative to a 12-hour moving average of the open price—with a 400\% volume threshold, supplemented by the 20-day EWMA and volatility conditions. This configuration effectively minimizes noise while preserving the model’s ability to detect true pump-and-dump activities, thus offering a reliable framework for real-world anomaly detection in cryptocurrency markets.

\bibliography{pump_biblio}

\end{document}